%% file: syn-asy.tex
\documentclass{eptcs}

\usepackage{makeidx}  
\usepackage{mathtools}
\usepackage{amsmath}
\usepackage{amssymb}
\usepackage{amsthm}
\usepackage{centernot}
\usepackage{mathpartir}
\usepackage{listings}
\usepackage{fancyvrb}
\usepackage{float}
\usepackage{stmaryrd
  ,latexsym,
  fontenc}
\usepackage{semantic}
\usepackage{graphicx}
\usepackage{caption}

\usepackage{tikz}
\usetikzlibrary{matrix,shapes,arrows}
\tikzstyle{block} = [rectangle, draw, fill=white!20, text width=4em, text centered, minimum height=3em]
\tikzstyle{line} = [draw, very thick, color=black!90, -latex']

\usepackage{actor}
\usepackage{logic}
\usepackage{shorthand}
\usepackage{translation}
\usepackage{lang}
\usepackage{RV}
\usepackage{booktabs}

\graphicspath{{FIGURES/}}

\newcommand{\elarva}{\textsf{eLarva}}

\newcommand{\detecter}{\textsf{detectEr}\xspace}
\newcommand{\qedd}{\hfill\ensuremath{\blacksquare}}

\theoremstyle{definition}
\newtheorem{example}{Example}[section]
\newtheorem{remark}{Remark}[section]

\newcommand{\toolFRM}{\texttt{FRM}}
\newcommand{\toolFls}{\texttt{ff}}
\newcommand{\toolNec}[2]{\texttt{[}#1\texttt{]}#2}
\newcommand{\toolAnd}[2]{#1\,\texttt{\&}\,#2}
\newcommand{\toolMax}[2]{\texttt{max(}#1\texttt{,}#2\texttt{)}}
 
\newcommand{\toolBool}[2]{\texttt{bool}\,#1\;\texttt{=>}\,#2} 

\newcommand{\bV}{\ensuremath{b}}
\newcommand{\bVV}{\ensuremath{c}}

\newcommand{\toolActOut}[2]{#1\,\texttt{!}\,#2}
\newcommand{\toolActIn}[2]{#1\,\texttt{?}\,#2}
\newcommand{\toolActCall}[2]{\texttt{call(}#1,#2\texttt{)}}
\newcommand{\toolActRet}[2]{\texttt{ret(}#1,#2\texttt{)}}
\newcommand{\toolTup}[1]{\texttt{\{}#1\texttt{\}}}

\newcommand{\toolN}[1]{\texttt{#1}}

\begin{document}

\title{On Synchronous and Asynchronous Monitor Instrumentation for Actor-based systems}
\author{Ian Cassar\footnote{The research work disclosed in this publication is partially funded by the \textit{Master it!} Scholarship Scheme (Malta).}
\institute{Computer Science, ICT, University of Malta.}
\email{ian.cassar.10@um.edu.mt}
\and
Adrian Francalanza
\institute{Computer Science, ICT, University of Malta.}
\email{adrian.francalanza@um.edu.mt}
}
\def\titlerunning{On Synchronous and Asynchronous Monitor Instrumentation}
\def\authorrunning{A. Francalanza \& I. Cassar}

\maketitle              

\begin{abstract}
We study the impact of synchronous and asynchronous monitoring instrumentation on runtime overheads in the context of a runtime verification framework for 
actor-based systems.   We show that, in such a context, asynchronous monitoring incurs substantially lower overhead costs. We also show how, for certain  properties that require synchronous monitoring, a hybrid approach can be used that ensures timely violation detections for the important events while, at the same time, incurring lower overhead costs that are closer to those of an asynchronous instrumentation. 
\end{abstract}

\section{Introduction}
\label{sec:introduction}
\input{intro.tex}

\section{Erlang Actors and the Yaws Webserver}
\label{sec:actor-systems-erlang}
\input{erlang.tex}

\section{The Logic}
\label{sec:language}

\input{language.tex}


\section{Introducing Synchronous Instrumentation}
\label{sec:synchr-instr}
\input{synchmon.tex}

\section{Hybrid Instrumentation}
\label{sec:hybr-instr}
\input{hybrid.tex}

\section{Conclusion}
\label{sec:conclusion}
\input{conc.tex}

\bibliographystyle{eptcs}
\bibliography{references}



\end{document}

%% file: intro.tex


Formally ensuring the correctness of  component-based, concurrent systems is an arduous task, mainly because exhaustive methods such as model-checking quickly run into state-explosion problems; this is typically caused by the 
multiple thread interleavings of the system being analysed, and the range of data the system can input and react to.    Runtime Verification (RV) \cite{Leu:RV:Overv} is an appealing compromise towards ensuring correctness, as it circumvents such scalability issues by verifying only the \emph{current} system execution.  Runtime Enforcement (RE)  \cite{FalconeFM12} builds on  RV  by automating recovery procedures once a 
violation is detected so as to mitigate or rectify the effects of the violation.  Together, these runtime techniques can be used as a disciplined methodology for augmenting systems with self-adaptation functionality.

Most RV and RE frameworks work by synthesising \emph{monitors} from properties specified in terms of a formal language, and then execute these monitors in tandem with the system. Online\footnote{By contrast, \emph{offline} monitoring typically works on \emph{complete} execution traces, and occasionally going  back and forth along this trace during analysis.\cite{SyncVSAsync:Rosu:2005} } monitoring, \ie the runtime analysis of a system from the \emph{partial} execution trace generated thus far, usually comes in two 
\emph{instrumentation} flavours.  In \emph{synchronous} runtime monitoring, system trace events are forwarded to the monitor while the system \emph{is paused}, waiting for the monitor to acknowledge back before it can continue executing (until the next trace event is generated). By contrast, in \emph{asynchronous} monitoring, the system \emph{does not pause} when trace events are generated;  instead, events are kept in a buffer and processed by the monitor at some later stage, 
thereby \emph{decoupling} the execution of the system from that of the monitor.     

Both forms of instrumentation have their merits. Synchronous monitoring guarantees \emph{timely detection} of property violations/satisfaction since the system and monitor execute in lockstep; this facilitates the \emph{runtime enforcement} of properties, where remedial action can be promptly applied to a system waiting for a monitor response. Although asynchronous monitoring may lead to late detections, it is \emph{less intrusive} than its synchronous counterpart.  Its instrumentation is easier to carry out and  does not necessarily require access to the system code prior to execution. The associated \emph{overheads} are also assumed to be \emph{lower} than its synchronous counterpart, and it is more of a \emph{natural fit} for settings with inherent notions of asynchrony, such as in 
distributed systems.   Asynchronous monitoring also poses a lower risk of compromising system behaviour   in the eventual case of an erroneous monitoring algorithm that forgets to acknowledge back or diverges (\ie enters an infinite internal loop) since, in asynchronous monitoring the system execution is decoupled from that of the monitor.

Issues relating monitor instrumentation are particularly relevant to actor-based \cite{Agha98afoundation,Mason99actorlanguages,actorsPaper} component systems.  Synthesising asynchronous monitors as actors is in accordance with the actor model of computation, requiring independent computing entities to execute in decoupled fashion so as to permit scalable coding techniques such as fail-fast design patterns \cite{Cesarini:2009}; such code organisations (using monitor actors called \emph{supervisors})  are prevalent in actor based languages and technologies such as Erlang \cite{Armstrong07,Cesarini:2009}, Scala \cite{actorsinscala} and AKKA \cite{akka}. However, there are cases where tighter analyses through synchronous monitoring may be required, particularly when timely detections improve the effectiveness of subsequent recovery procedures.  Crucially, the appropriate monitor instrumentation needs also to incur low runtime overheads for it to be viable.    

In this paper we investigate issues related to monitor instrumentation in actor-based component systems  constructed using Erlang, a mature, industry-strength language used to build fault-tolerant, self-adaptive systems with high degrees of resilience \cite{Armstrong07}.  As a representative Erlang system for our experiments  we consider Yaws \cite{yaws:11,yaws:12}, a high performance HTTP webserver that makes extensive use of actors to handle multiple concurrent client connections.   In order to show how this study can be used to verify and enforce correct behaviour of Erlang systems, we also employ \detecter \cite{Fra:Sey:13,FraSeyTool13}, a component-based RV tool designed to monitor for Erlang safety properties.  Within this setting: 
\begin{enumerate}
\item  we \emph{design and implement }mechanisms for \emph{synchronous instrumentation}; asynchronous monitoring is natively supported through mechanisms offered by the Erlang Virtual Machine (EVM) --- these are presently employed by \detecter, but also other tools such as \cite{elarva:2012,Exago}.
\item we  \emph{quantify} the relative computational overhead incurred by synchronous and asynchronous monitoring; it is generally accepted that asynchronous monitoring incurs less overheads, but we are not aware of any studies that attempt to quantify by how much.
\item we \emph{devise} a novel \emph{hybrid instrumentation technique} that guarantees timely detections while incurring overheads comparable to those of asynchronous monitoring.  We also \emph{asses} the effectiveness of the technique \wrt the other methods of instrumentation.
\item we \emph{integrate} the different modes of instrumentation within  \detecter,  allowing one to specify correctness properties with multiple violation conditions where some violations are monitored asynchronously, some are monitored  synchronously, and for others monitoring switches between synchronous and asynchronous modes. Although tools for synchronous and asynchronous monitoring exists, we are not aware of any that combine the two within the same monitor execution.  
\end{enumerate}
Although the work in this paper focusses on detection, it lays the necessary foundations for implementing effective enforcement mechanisms that are launched as soon as violations are detected, allowing us to augments systems with self-adaptive functionality.

The  rest of the
paper is structured as follows. \secref{sec:actor-systems-erlang} briefly outlines actors in Erlang and introduces Yaws. \secref{sec:language} describes the logic used by the tool \detecter and shows how this logic can specify safety properties for Yaws. In \secref{sec:synchr-instr}, we devise a technique for instrumenting synchronous monitors for an actor system and asses the relative overhead costs compared to the existing asynchronous instrumentation. \secref{sec:hybr-instr} presents a novel hybrid setup where one can specify which events are to be monitored synchronously or asynchronously. We asses the overheads incurred by the new instrumentation technique in \secref{sec:evaluation}. 
\secref{sec:conclusion} concludes by outlining future and related work.

%% file: erlang.tex
Erlang \cite{Armstrong07} is an actor-based \cite{actorsPaper}  programming language where \emph{processes} (\ie actors) are threads of execution that are uniquely identified by a \emph{process identifier} and own their own \emph{local} memory.  Erlang processes execute \emph{asynchronously} to one another, interacting through \emph{asynchronous messages}: instead of sharing data, processes explicitly send a \emph{copy} of this data to the destination process (using the unique identifier as address); messages are received at a process \emph{mailbox} (a form of message buffer) and can be exclusively read at a later stage by the process owning the mailbox. Processes may also \emph{spawn} other processes at runtime, and communicate locally-known (\ie private) process identifiers as messages.  Implementation wise, Erlang processes are relatively lightweight, and language coding practices recommend the use of concurrent processes wherever possible.  These factors ultimately lead to systems made up of independently-executing components 
that are more scalable, maintainable, and use the underlying parallel architectures more efficiently \cite{Cesarini:2009}.   

At its core, Erlang is \emph{dynamically-typed}, and function calls with mismatching parameters are typically detected at runtime.  Function definitions are named and organised in uniquely-named modules. Within a particular module, there can be more than one function with the same name, as long as these names have different arities (\ie number of parameters).  Thus, 
every function is uniquely identified by the triple \emph{module\_name:function\_name:arity}. 

Erlang offers a number of fault-tolerance mechanisms. Since asynchronous  messages  may  reach a mailbox 
in a different order than the one intended, the (mailbox) read construct uses \emph{pattern matching} to allow a process to retrieve the first message in the mailbox matching a pattern, possibly reading messages out of order. 
Erlang also offers a mechanism for \emph{process linking}, whereby a process receives a notification message in its mailbox when a linked process fails \ie terminates abnormally.  Erlang systems are often structured by the \emph{supervisor} pattern which is built on linking: processes at the system fringes are encouraged to \emph{fail-fast} when an error occurs (as opposed to handling the error locally), leaving error handling to linked supervisor processes \cite{Cesarini:2009}.

\subsection{YAWS: A webserver written in Erlang}
\label{sec:case-study}

Yaws\cite{yaws:11,yaws:12} is a high-performance, component-based HTTP webserver written in Erlang.  For every client connection, the server assigns a dedicated (concurrent) handler that 
services HTTP client requests, thereby parallelising processing for multiple clients.  
Its implementation 
relies on the lightweight nature of  Erlang processes to 
efficiently handle a vast number of simultaneous client connections.

\begin{figure}[t]
  \centering
  \begin{tikzpicture}[>=latex,auto,thick]
    \begin{scope}[draw=blue!50,fill=blue!20,minimum size=0.8cm]
      \node (client) at ( 0,3) [shape=rectangle,draw,fill] {Client};
      \node (curHandler) at ( 5,3) [shape=rectangle,draw,fill] {\emph{Current} Handler};
      \node (acceptor) at ( 10,3) [shape=rectangle,draw,fill] {Acceptor};       
    \end{scope}
   \begin{scope}[draw=black,dashed,minimum size=0.8cm]
    \node (newHandler) at ( 5,1.5) [shape=rectangle,draw] {\emph{New} Handler};
   \end{scope}
    \begin{scope}[draw=red,fill=red]

      \draw[->] (client) [bend left=15] to node [above]{(1) connect} (curHandler); 
      \draw[->] (curHandler) [bend left=15] to node [above]
      {(2) \{\textsl{HandlerPid}, \toolN{next}, \textsl{ClientPort}\}} (acceptor);
      \draw[->] (client) [bend right=15] to node  [below]{(4) HTTP requests} (curHandler); 
     \end{scope}
    \begin{scope}[draw=black, dashed]
      \draw[->] (acceptor) to node{(3)spawn} (newHandler);
    \end{scope}
  \end{tikzpicture}
  \caption{Yaws client connection protocol}
\label{fig:yaws-protocol}
\end{figure}
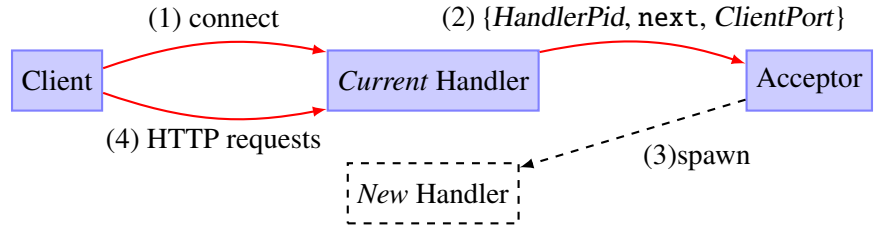

The Yaws protocol for establishing client connections is depicted in \figref{fig:yaws-protocol}. This protocol 
uses an \emph{acceptor} component which, upon creation, spawns a \emph{connection handler} to be assigned to the next client connection. Subsequently, the acceptor blocks waiting for messages in its mailbox, while the unassigned handler waits for the next TCP connection request. Clients send  connection requests through standard TCP ports $(1)$, which are received  as messages in the \emph{handler}'s mailbox.   The current handler accepts these requests by reading the \resp message from its mailbox and $(2)$  sending a message containing its own \emph{pid} and the \emph{port} of the connected client to the \emph{acceptor}; this acts as a notification that it is now engaged in handling the connection of a specific client. Upon receiving the message, the \emph{acceptor} unblocks, records the information sent by the handler for supervision purposes (\eg restarting the handler in case it crashes) and $(3)$ spawns a \emph{new} handler  listening for future  connection requests. 

Once it is assigned a handler, the connected client then engages \emph{directly} with it using $(4)$ standard HTTP requests; these normally consist of six (or more) HTTP headers containing the information such as the client's User Agent, Accept-Encoding and the Keep-Alive flag status. 
HTTP request information is \emph{not} sent in one go but follows a protocol of messages: it starts by sending the \toolN{http\_req}, followed by six \toolN{http\_header} messages containing client information, terminated by a final \toolN{http\_eoh} message. The dedicated connection handler inspects the request 
and services the \resp HTTP request 
accordingly.

%% file: language.tex
We conduct our investigations using 
the component-based runtime verification tool called \detecter. In \cite{Fra:Sey:13}, the authors present a tool that synthesises concurrent  monitors (as systems of Erlang processes) from 
a syntactic subset of the modal $\mu$-calculus specifying \emph{safety} properties for Erlang systems; the sublogic is called \SHML \cite{aceto:SHML}. In 
\cite{Fra:Sey:13}, these monitors \emph{asynchronously} analyse the system so as to verify for runtime violations of the respective formulas.  Actor based systems such as those constructed using Erlang typically grown and shrink in size as computation progresses.\footnote{For instance, the Yaws webserver of \secref{sec:case-study} creates a new handler component for every new client request.}  Accordingly, the component-based monitors generated by \detecter\ are able to scale with the current size of the system being monitored.   

\begin{figure}[t]
  \centering
  \begin{align*}
    \hV,\hVV \in \toolFRM & \;\bnfdef\; \toolFls \;\bnfsepp\; \toolAnd{\hV}{\hVV} \;\bnfsepp\; \toolNec{\actE}{\hV}  \;\bnfsepp\; \hVarX \;\bnfsepp\; \toolMax{\hVarX}{\hV} \;\bnfsepp\;  \toolBool{\bV}{\hV} 
  \end{align*}
  \begin{align*}
    \hmeaning{\toolFls} & \deftxt \emptyset\\
    \hmeaning{\toolAnd{\hV}{\hVV}} & \deftxt \hmeaning{\hV} \cap \hmeaning{\hVV}\\
    \hmeaning{ \toolNec{\actE}{\hV}} & \deftxt \sset{\actV \;|\;\; \Bigl(\actV \wtraS{\actEE} \actVV  \text{ and } \match(\actE,\actEE) = \sigma\Bigr) \;\text{ implies }\;\actVV \in \hmeaning{\hV\sigma} }\\
    \hmeaning{\toolMax{\hVarX}{\hV}} & \deftxt \bigcup \sset{S \;|\; \; S \subseteq \hmeaning{\hV\sset{\hVarX\mapsto S}}}\\
    \hmeaning{\toolBool{\bV}{\hV}}  & \deftxt
    \begin{cases}
      \hmeaning{\hV} & \text{if } \bV\Downarrow \btt\\
      \Actors & \text{if } \bV\Downarrow \bff
    \end{cases}
  \end{align*}
  \caption{The Logic and its Semantics}
  \label{fig:logic}
\end{figure}

The syntax of our logic is defined inductively using the BNF description in \figref{fig:logic}.  The logic is an extension to that of \cite{Fra:Sey:13}, facilitating the expression of properties dealing with data.   It is parametrised by a set of boolean expressions, $\bV,\bVV\in\Bool$, equipped with a \emph{decidable} evaluation function, $\bV\Downarrow\vV$ where $\vV \in \sset{\btt,\bff}$, and a set of actions $\actE,\actEE\in\Act$ that may universally quantify over data values. 
It assumes two distinct denumerable sets of \emph{term variables}, $\xV,\xVV,\ldots\in\Vars$, used in actions  and boolean expressions,  and \emph{formula variables} $\hVarX,\hVarY,\ldots\in\LVars$, used to define recursive logical formulas; in what follows, we work up-to $\alpha$-conversion of bound variables.\footnote{\detecter\ renames duplicate variables  accordingly during a pre-processing phase.} Formulas include falsity, \toolFls, conjunctions,  \toolAnd{\hV}{\hVV}, modal necessities, \toolNec{\actE}{\hV}, and maximal fixpoints \toolMax{\hVarX}{\hV} from \cite{Fra:Sey:13}.  One important extension to \cite{Fra:Sey:13} is that  actions, \actE, used in necessity formulas, \toolNec{\actE}{\hV}, may contain term variables that \emph{pattern-match} with actual (closed) actions. We also extend the logic with boolean guards, \toolBool{\bV}{\hV}, that may contain term variables introduced by necessity formulas.  A formula \toolNec{\actE}{\hV} is thus a \emph{binder} for the variables used in \actE\ in the subformula \hV; similarly \toolMax{\hVarX}{\hV} is a binder for \hVarX\ in \hV.  To improve readability, we sometimes denote term variables introduced by \actE\ in \toolNec{\actE}{\hV}  that are not used in the subformula  \hV\ as underscores, $\_$.

The semantics of the logic is defined over an arbitrary labelled transition system (LTS), $\langle \Actors, \Act\cup\sset{\tau}, \rightarrow \rangle$, where $\actV,\actVV\in\Actors$ is the set of nodes denoting actor systems,  $\Act\cup\sset{\tau}$ is a set of actions including a silent (internal) action $\tau$, and $\rightarrow$ is a ternary relation of type $\Actors\times(\Act\cup\sset{\tau}) \times \Actors$.  In \cite{Fra:Sey:13}, the authors show how Erlang programs can be given an LTS  semantics and, in this paper, we conveniently adopt that semantics.  In practice, however, other LTS semantics for the language, such as \cite{fredlund:unifiedsemantics,fredlund:phd} can be used instead.   We write $\actV \traS{\actE} \actVV$ (\resp  $\actV \traS{\tau} \actVV$) in lieu of $(\actV,\actE,\actVV) \in \rightarrow$ (\resp $(\actV,\tau,\actVV) \in \rightarrow$) and write $\actV\wtraS{\actE}\actVV$ to denote $\actV(\traS{\tau})^\ast\cdot\traS{\actE}\cdot(\traS{\tau})^\ast\actVV$.  

Our logic semantics is presented in \figref{fig:logic}, through the denotational function $\hmeaning{-}:: \toolFRM \rightarrow \pset{\Actors}$, defined inductively on the structure of the formula.   The definition assumes \emph{well-formed} formulas \ie formulas where all variables are \emph{bound} and formula variables are \emph{guarded} (appear under a necessity formula). 
We say \actV \emph{satisfies} \hV, denoted as $\actV\models\hV$, whenever $\actV\in\hmeaning{\hV}$.    The semantics follows that of \cite{Fra:Sey:13}. No actor system satisfies  \toolFls, whereas actors satisfying \toolAnd{\hV}{\hVV} must satisfy both \hV\ and \hVV. In our extension, the necessity formula is imbued with \emph{pattern-matching} functionality, represented by the function $\match(\alpha,\beta)$ matching a (possibly) open action $\alpha$ (\ie with term variables in it) with a closed action $\beta$, returning a substitution, $\sigma:: \Vars\rightharpoonup \Val$, whenever successful.      
\begin{align}
  \label{eq:1}
  \match(\,\toolActOut{\toolN{server}}{\toolTup{x,\toolN{ack},y}},\, & \toolActOut{\toolN{server}}{\toolTup{5,\toolN{ack},\toolN{joe}}} \,) &&\; =\; \sset{x\mapsto 5, y\mapsto \toolN{joe}} \\
  \label{eq:2}
  \match(\,\toolActOut{\toolN{server}}{\toolTup{5,\toolN{ack},\toolN{joe}}},\,& \toolActOut{\toolN{server}}{\toolTup{5,\toolN{ack},\toolN{joe}}} \,) &&\; =\; \sset{} \\
  \label{eq:3}
  \match(\,\toolActOut{\toolN{client}}{\toolTup{x,\toolN{ack},y}},\,& \toolActOut{\toolN{server}}{\toolTup{5,\toolN{ack},\toolN{joe}}} \,)  &&\text{ is undefined} \\
  \label{eq:4}
  \match(\,\toolActIn{\toolN{server}}{\toolTup{x,\toolN{ack},y}},\,& \toolActOut{\toolN{server}}{\toolTup{5,\toolN{ack},\toolN{joe}}} \,)  &&\text{ is undefined}
\end{align}
For example,  in (\ref{eq:1}) the open output action $\toolActOut{\toolN{server}}{\toolTup{x,\toolN{ack},y}}$ is successfully matched with the output action \toolActOut{\toolN{server}}{\toolTup{5,\toolN{ack},\toolN{joe}}}, where $x$ and $y$ are pattern matched with the values $5$ and \toolN{joe} \resp In (\ref{eq:2}) the two closed actions are matched (exactly), returning the empty substitution. The mismatch in (\ref{eq:3})  is due to mismatching destinations of the output actions \ie \toolN{server} versus \toolN{client}, whereas the mismatch in (\ref{eq:4}) is because the input action $\toolActIn{\toolN{server}}{\toolTup{x,\toolN{ack},y}}$ cannot be pattern matched with actions of a different kind (\eg output).  Necessity formulas \toolNec{\actE}{\hV} are satisfied by all actor systems \actV observing the condition that, \emph{whenever}  pattern-matchable actions   $\actEE$ are performed (yielding substitution $\sigma$), the resulting actors \actVV that are transitioned to \emph{must} satisfy $\hV\sigma$. Note that actors that \emph{do not} perform any pattern-matchable actions trivially satisfy \toolNec{\actE}{\hV}.  Formula \toolMax{\hVarX}{\hV} denotes the maximal fixpoint of the functional \hmeaning{\hV}; following standard fixpoint theory \cite{tarski:55}, this is characterised as the union of all post-fixpoints $S\in\pset{\Actors}$.  Guarded formulas \toolBool{\bV}{\hV} equate to \hV\ the whenever \bV\ evaluates to \btt\ but are trivially satisfied whenever  \bV\ evaluates to \bff.  

\begin{example} \label{ex:simple-formula}  Consider the formula
 \begin{equation} \label{eq:7:language}
    \toolMax{\hVarX}{\toolNec{\toolActIn{\toolN{server}}{\toolTup{\toolN{succ},x,y}}}{\;\toolNec{\toolActOut{y}{z}}{\;
          \left(
            {(\toolBool{(z\!=\!x\!+\!1)}{\hVarX\,})}\; 
            \toolAnd{}{\;(\toolBool{(z\neq x\!+\!1)}{\toolFls\,})}
          \right)
        }}}
\end{equation}
It states that a satisfying system repeatedly observes the condition, \ie \toolMax{\hVarX}{\ldots}, that whenever it accepts an input at actor \toolN{server} with values matching the pattern $\toolTup{\toolN{succ},x,y}$ --- denoting a server request from client $y$ for a successor computation, the \toolN{succ} message tag,  for value $x$ --- followed by a reply output message sent to $y$ with the answer $z$, then the answer is indeed an increment on value $x$.  \qedd
\end{example}

\begin{remark}
  Apart from input and output actions from the original tool presentation of \cite{Fra:Sey:13}, our
  logic extension also considers actions denoting function calls and returns,
  $\toolActCall{\textsl{Pid}}{\toolTup{\textsl{module},  
      \textsl{function},\textsl{values}}}$ and
  $\toolActRet{\textsl{Pid}}{\toolTup{\textsl{module},
      \textsl{function},\textsl{arity},\textsl{values}}}$ \resp  These are
  needed in actual Erlang implementations because certain output and input actions may be
  abstracted away inside the function calls of system libraries,
  making them (directly) unobservable to the instrumentation mechanism.
  However, there are cases where we can still observe these actions
  (and the data associated with them) \emph{indirectly}, through the
  calls and returns of the functions that abstract them.
\end{remark}

\subsection{Monitoring for safety properties in Yaws}
\label{sec:monit-safety-prop}


The logic is expressive enough to express a number of safety properties for Yaws, including the Directory Traversal Vulnerability found in earlier versions of the software and reported on the reputable exploit-db website \cite{yaws-exploit} 
. We here discuss a second safety property for the webserver.

If we assume the existence of a (decidable) predicate called $\textsl{isMalicious}()$, which can determine whether the client will engage in security-breaching activity from the 6 HTTP headers sent to the handler, we can use the logic of \figref{fig:logic} 
to specify the safety property stated in (\ref{eq:5}).  The property requires that, for \emph{every} client connection established (determined from message $(2)$ of \figref{fig:yaws-protocol} and denoted in property (\ref{eq:5}) as the action $\toolActOut{\toolN{AcceptorPid}}{\toolTup{handID,\toolN{next},\_}}$) the following subproperty must hold: for \emph{every} HTTP request, the respective headers communicated ($h1$ to $h6$) do not amount to a potentially security-breaching request (as determined by the predicate $\textsl{isMalicious}()$).
\begin{equation}
\begin{array}{l}
    \toolMax{\hVarX}{\\\quad\toolNec{\toolActOut{\toolN{AcceptorPid}}{\toolTup{handID,\toolN{next},\_}}}\\
      \quad\toolAnd{\Bigl( X\;}{}\\\
      \quad\;\toolMax{\hVarY}{\toolNec{\toolActRet{handID}{\toolTup{\toolN{yaws}, \toolN{do\_recv},\toolN{3},\toolTup{\toolN{ok},\toolTup{\toolN{http\_req},\toolN{GET},\_ ,\_}}}}}\\
    \qquad\toolNec{\toolActRet{handID}{\toolTup{\toolN{yaws}, \toolN{do\_recv},\toolN{3},\toolTup{\toolN{ok},h1}}}}
    \toolNec{\toolActRet{handID}{\toolTup{\toolN{yaws}, \toolN{do\_recv},\toolN{3},\toolTup{\toolN{ok},h2}}}}\\
    \qquad\toolNec{\toolActRet{handID}{\toolTup{\toolN{yaws},\toolN{do\_recv},\toolN{3},\toolTup{\toolN{ok},h3}}}}
    \toolNec{\toolActRet{handID}{\toolTup{\toolN{yaws},\toolN{do\_recv},\toolN{3},\toolTup{\toolN{ok},h4}}}}\\
    \qquad\toolNec{\toolActRet{handID}{\toolTup{\toolN{yaws},\toolN{do\_recv},\toolN{3},\toolTup{\toolN{ok},h5}}}}
    \toolNec{\toolActRet{handID}{\toolTup{\toolN{yaws},\toolN{do\_recv},\toolN{3},\toolTup{\toolN{ok},h6}}}}\\
    \qquad
            \left(\begin{array}{l}
            (\toolBool{(\textsl{isMalicious}(h1,h2,h3,h4,h5,h6))}{\toolFls})\\
            \toolAnd{}{\bigl(\toolBool{(\lnot \textsl{isMalicious}(h1,h2,h3,h4,h5,h6))}{}\\
            \qquad\qquad
            \toolNec{\toolActRet{handID}{\toolTup{\toolN{yaws},\toolN{do\_recv},\toolN{3},\toolTup{\toolN{ok},\toolN{http\_eoh}}}}}{\hVarY})}\bigr)
          \end{array}
          \right)\\
     \quad\Bigr) }}
\end{array}\label{eq:5}
\end{equation}
The logical formula stated in \eqref{eq:5} specifies this property  by using two nested recursive formulas.  The outer one, $\toolMax{\hVarX}{\ldots}$, refers to each assigned handler by pattern-matching with the term variable \emph{handID}, whereas the inner one, $\toolMax{\hVarY}{\ldots}$, uses this value to pattern match with the header term variables, \emph{h1} to \emph{h6}, for every iteration of the HTTP request protocol; boolean guarded formulas are then used to determine whether these HTTP requests violate the property or not.  We note  that,  whereas the handler messages to the acceptor  are observed \emph{directly} (\ie the output action in the outer recursive formula), the client HTTP messages received by the handler have to be observed \emph{indirectly} through the return values (of the form $\toolTup{\toolN{ok},\_}$) of the invoked function \toolN{do\_recv}.\footnote{Defined in module \toolN{yaws} with arity \toolN{3}.}   Instrumentation allowing a direct observation of these  actions is complicated by the fact that the client TCP messages are sent through functions from the \toolN{inet} Erlang library, which is part of the Erlang Virtual Machine kernel.

\subsection{Component-based Monitor Generation}
\label{sec:monit-synth-scal}

\begin{figure}[t]
  \centering
  \begin{tikzpicture}[>=latex,auto,thick]
    \begin{scope}[draw=blue!50,fill=blue!20,minimum size=0.8cm]
      \node (client2) at ( 0,4.5) [shape=rectangle,draw,fill] {Client};
      \node (cHandler2) at ( 4,4.5) [shape=rectangle,draw,fill] {Handler};
      \node (mon2) at (7,4.5)  [shape=rectangle,draw] {Monitor};
      
      \node (client) at ( 0,3) [shape=rectangle,draw,fill] {Client};
      \node (cHandler) at ( 4,3) [shape=rectangle,draw,fill] {Handler};
      \node (mon) at (7,3)  [shape=rectangle,draw] {Monitor};
      
      \node (acceptor) at ( 11,3) [shape=rectangle,draw,fill] {Acceptor};  
     \node (newHandler) at ( 4,1.5) [shape=rectangle,draw,fill] {(free) Handler};
     \node (newmon) at (7,1.5)  [shape=rectangle,draw] {Monitor};     
    \end{scope}
   \begin{scope}[draw=black,dashed,minimum size=0.8cm] 
   \end{scope}
    \begin{scope}[draw=red,fill=red]
      \draw[<->] (client) to node{assigned} (cHandler);
      \draw[<->] (client2) to node{assigned} (cHandler2); 
     \end{scope}
    \begin{scope}[draw=black, dashed]
      \draw[<-] (mon) to node{trace} (cHandler);
      \draw[<-] (mon2) to node{trace} (cHandler2);
      \draw[<-] (newmon) to node{trace} (newHandler);
    \end{scope}
  \end{tikzpicture}
  \caption{Component-based Monitoring}
\label{fig:yaws-monit}
\end{figure}
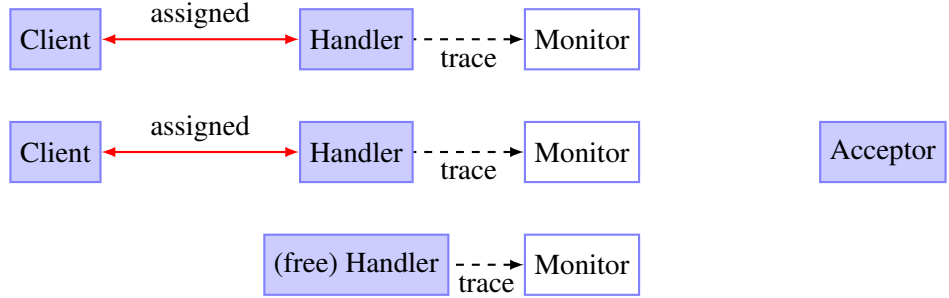

The monitors generated by \detecter are \emph{not} monolithic, but consist of systems of concurrent (sub)monitors, each analysing different parts of a system under examination.  For instance, in the case of property \eqref{eq:5}, \detecter first generates a submonitor that listens for the action $\toolActOut{\toolN{AcceptorPid}}{\toolTup{handID,\toolN{next},\_}}$  from the unassigned (free) handler waiting on the TCP port.  Once this action is detected, it spawns a new submonitor to listen for the new handler, while it continues monitoring for the handler that is now assigned to a client.  Thus, after two client connections are accepted, we end up with the configuration depicted in \figref{fig:yaws-monit} (see \cite{Fra:Sey:13} for details).  Note that the submonitors need not communicate with 
each other since a violation detected by one submonitor immediately means that the global property is violated.  

%% file: synchmon.tex
\newcommand{\madv}{\texttt{advices:}\xspace}
\newcommand{\aadv}{\texttt{after\_advice}\xspace}
\newcommand{\badv}{\texttt{before\_advice}\xspace}
\newcommand{\uadv}{\texttt{upon\_advice}\xspace}


In \cite{Fra:Sey:13}, the monitors generated by \detecter are exclusively asynchronous, using the tracing mechanism offered by the EVM OTP platform libraries \cite{ErlangOTP} which do not require instrumentation at source-code level; instead VM directives generate trace messages for specified execution events that are sent to a specially-designated \emph{tracer} actor.   However, as was argued in \secref{sec:introduction}, there are various cases where safety properties may need to be monitored synchronously.   

There are a number of ways how one can layer synchronous monitoring atop of an asynchronous computational model. For instance, one can insert actual monitoring functionality \emph{within} the sequential thread execution of each actor (in the style of \cite{EfficientDec,chen-rosu-2007-oopsla,polyLarva-CFMP12}) and then have the monitoring code (scattered across independently executing actors) synchronise, as required, in a \emph{choreographed} setup \cite{EfficientDec,FGP12DistribRV}.   Instead, we opt for an \emph{orchestrated} solution, whereby individually monitored actors are only instrumented to report monitored actions to a (conceptually) centralised monitoring setup that receives all reported actions and performs the necessary checking.\footnote{Although conceptually centralised, the orchestrator monitor consists of independent, concurrent sub-monitors as discussed in \secref{sec:monit-synth-scal}.}  There are a number of reasons for choosing such a setup.  First, the instrumentation code at the system side is minimal, leaving the instrumented code close to the original. Moreover, monitoring  is \emph{consolidated} into a group of concurrent actors that are separate from the monitored system,  improving manageability (\eg parts of the system may crash leaving the monitor (largely) unaffected).  More importantly, it allows us to perform a like-with-like comparison with the existing asynchronous setup present in \detecter, thus obtaining a more precise comparison between the relative overheads of synchronous and asynchronous monitoring.


\begin{figure}[t]
  \centering
  \begin{tikzpicture}[>=latex,auto,thick]
    \begin{scope}[draw=blue!50,fill=blue!20,minimum size=0.8cm]
      \node (sys) at ( -0.5,1) [shape=rectangle,draw,fill,text width=4cm, minimum height = 3cm, text centered]      
          {...\\ Event e1 occurs;\\ Get data d1 from e1;\\ Block on nonce(e1);\\...\\Event e2 occurs;\\ Get data d2 from e2;\\Block on nonce(e2);\\...};
       \node (mon) at ( 8.55,1) [shape=rectangle,draw,fill,text width=6.1cm, minimum height = 3cm]
       {loop(Nonce)\;$\rightarrow$\\ $\quad$send\_ack(Nonce), \\ $\quad$\{EVT,Nonce2\} = recv\_event(), \\ $\quad$HasPatternMatched = handle(EVT), \\ $\quad$if(HasPatternMatched) $\rightarrow$ \\ $\qquad$loop(Nonce2); \\ $\quad$else $\rightarrow$\\ $\qquad$ send\_ack(Nonce2)\\ $\quad$end. };  
        \node (sysN) at (-0.5,3.5) {\textbf{System}}; 
        \node (sysN) at (9,3.5) {\textbf{Monitor}}; 
    \end{scope}
     \node (sys0) at (1.5,2.3){};
       \node (mon0) at (5.5,2.3){};
       \node (sys1) at (1.5,1.5){};
       \node (mon1) at (5.5,1.5){};
       \node (sys2) at (1.5,1.2){};
       \node (mon2) at (5.5,1.2){};
       \node (sys3) at (1.5,0.1){};
       \node (mon3) at (5.5,0.1){};
       \node (sys4) at (1.5,-0.2){};
       \node (mon4) at (5.5,-0.2){};
    \begin{scope}[draw=red,fill=red, dashed]
    	\draw[<-] (sys0)  to node{\textbf{ack}(init\_nonce)} (mon0); 
      \draw[->] (sys1)  to node{\textbf{event}(d1,nonce1)} (mon1); 
      \draw[->] (mon2)  to node{\textbf{ack}(nonce1)} (sys2); 
      \draw[->] (sys3) to node{\textbf{event}(d2,nonce2)} (mon3); 
      \draw[->] (mon4) to node{\textbf{ack}(nonce2)} (sys4); 
    \end{scope}
  \end{tikzpicture}
  \caption{A high level description of  the synchronous event monitoring protocol.}
       \label{fig:sync_protocol}
\end{figure}
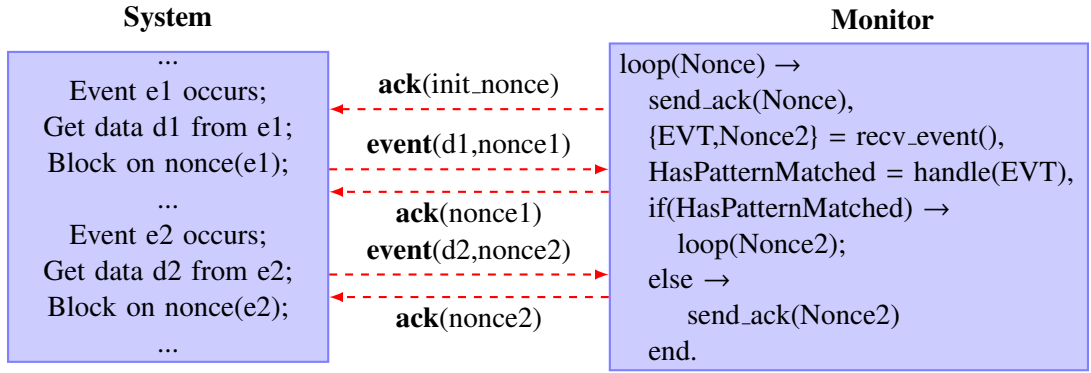



\figref{fig:sync_protocol} depicts the synchronisation protocol between the system instrumented code and the monitor for two monitored events; synchrony is achieved through handshakes over asynchronous messages between the two parties. The monitoring loop starts by sending an acknowledgement message to the system, signalling it to execute until it produces the next monitored event.  When this point is reached,
the instrumentation code  at the system side extracts the necessary data associated with the event, sends it as a trace message to the monitor and pauses by blocking on an unpause-acknowledgement message from the monitor.  Since, the (acknowledgement) asynchronous messages may get reordered in transit (potentially interfering with this protocol)  the instrumentation code 
generates a unique \emph{nonce} for every monitored event and sends it with the event data.  In return, the next time the monitor sends back an  acknowledgement message, it includes this unique nonce; this allows the instrumentation code to pattern-match for (and possibly read out-of-order) mailbox inputs containing this nonce, and unblock to the corresponding acknowledgement.    The monitoring loop outlined in \figref{fig:sync_protocol} corresponds to the pattern-matching functionality required by a necessity formula $\toolNec{\alpha}{\hV}$. For instance, if the pattern is not matched, it acknowledges immediately to the system to continue executing and terminates the monitoring \ie command \texttt{send\_ack(Nonce2)}. However, if the pattern matches, there is still a chance that a violation can be detected: it therefore delegates the acknowledgement (with the corresponding nonce) to the next part of the monitor (corresponding to $\hV$ in $\toolNec{\alpha}{\hV}$) \ie command \texttt{loop(Nonce2)}.  Crucially, if $\hV = \toolFls$, the monitor \emph{does not} send back the acknowledgement, blocking the offending system indefinitely while flagging the violation; in a runtime enforcement setup, the execution of a recovery procedure would substitute the violation flagging. 

\subsection{Instrumentation through Aspect-Oriented Weaving}
\label{sec:instr-thro-aspect}

The instrumentation was carried out by \emph{extending} an Aspect Oriented Framework for Erlang \cite{AOPErlang} 
to add the necessary instrumentation  in the system code; \cite{AOPErlang} did not support aspects for output and input events.  
Our aspect-based instrumentation uses a purpose built Erlang module called \texttt{advices.erl} containing the 3 type of advices used by our AOP injections.  For send and function call events, the AOP weaves \badv advices  reporting the event data to the monitor. For function returns, the AOP weaves \aadv advices
at the source location where the function is invoked; in this case, after advices are required since the return values are only known at that point.   The 
weaving concerns mailbox reading events, performed through the \texttt{recieve} construct, was not as straightforward.  Since a \texttt{recieve} may contain multiple pattern-matching guarded clauses 
\[\texttt{recieve } \textsl{guard}_1 \texttt{ -> } \textsl{expression}_1 \texttt{ ; } \ldots\texttt{ ; } \textsl{guard}_n \texttt{ -> } \textsl{expression}_n \texttt{ end}\] 
the AOP weaves \uadv advice for each guarded expression (\ie after each  \texttt{->}).  At runtime, only \emph{one} \texttt{recieve} guarded expression is triggered, at which point the necessary pattern-matched data of the event is known and can be reported to the monitor by the advice.

\subsection{Preliminary results}
\label{sec:preliminary-results}

 We conducted a number of experiments to asses the relative overheads between synchronous and asynchronous monitoring. We synthesised numerous properties such as \eqref{eq:5} for a Yaws server installation handling varying amounts of client connections and HTTP requests.  The graphs obtained in \figref{fig:combined_results} clearly show that synchronous monitoring incurs higher overheads than its asynchronous counterpart in terms of CPU Utilisation,  memory consumption and the latencies it introduces; \figref{fig:combined_results} shows substantial responsiveness degradation when handling typical loads of client requests.      To rule out any gains obtained through the efficiencies of the OTP tracing platform, we also created our own version of asynchronous monitoring that uses aspect orientation (but without blocking the system);  for certain measures, \eg memory consumption, the overhead discrepancies were even larger (see \figref{fig:combined_results}).  


%% file: hybrid.tex
\newcommand{\toolSNec}[2]{\texttt{[|}#1\texttt{|]}#2}
\newcommand{\sNonce}[1]{\textsl{nonce}(#1)}
\newcommand{\toolSFls}{\texttt{sff}\xspace}


Despite the benefits of synchronous monitoring, the associated overheads obtained from our preliminary results are substantially higher so as to make it infeasible in practice.  We therefore devise an alternative instrumentation strategy  with the aim of guaranteeing timely violation detections while incurring lower overheads that are closer to those incurred by asynchronous instrumentation.  The key insight is that, 
in order to attain timely detections, the instrumentation need not require the system to execute in lockstep with the monitor for \emph{every} monitored event leading to the violation. Instead,  (expensive) synchronous event monitoring can be limited to  the \emph{final} event \emph{preceeding} a violation, letting the system execute in decoupled fashion otherwise.   Intuitively, for the  logic of  \figref{fig:logic}  these final events 
and the 
necessity 
actions preceeding (directly or indirectly) a \toolFls\ forumla.  


\begin{example}\label{ex:hybrid-instrumentation}
  Recall property (\ref{eq:7:language}) from \exref{ex:simple-formula}.  In order to synchronously detect a violation in this property,
  only action \toolNec{\toolActOut{y\!}{\!z}}{} needs to be synchronously monitored, since it
  precedes a \toolFls\ forumla (interposed by a conjunction and a
  boolean guard).  Action
  $\toolNec{\toolActIn{\toolN{server}}{\toolTup{\toolN{succ},x,y}}}{}$
  can be  asynchronously monitored, without affecting the timeliness of detections. \qedd
\end{example}

\subsection{Logic Extensions}
\label{sec:logic-extensions}

We extend the syntax of the logic introduced in \secref{sec:language} by two constructs: a \emph{synchronous false} formula and a \emph{synchronous necessity} formula with a semantics analogous to  that of \toolFls\ and \toolNec{\actE}{\hV} \resp (see \figref{fig:logic})
\begin{align*}
    \hV,\hVV \in \toolFRM & \;\bnfdef\; \ldots \;\bnfsepp\; \toolSFls  \qquad \emph{(synchronous false)} \;\bnfsepp\;  \toolSNec{\actE}{\hV} \qquad \emph{(synchronous necessity)} 
\end{align*}
In the extended logic, formulas carry additional  instrumentation information relating to how they need to be runtime-monitored: by default, all the monitoring is asynchronous, unless one specifies that  a violation is to be synchronously monitored,  \toolSFls, of that a particular event needs to be synchronously monitored,  \toolSNec{\actE}{\hV}.\footnote{Synchronous event monitoring can be used  to engineer synchronisation points during periods where the system is not required to be immediately responsive at which point a monitor is allowed  to catch up with a system execution,  as in 
  \cite{CP12FF}.}

\begin{example} \label{ex:synch-asynch}  We can refine property~\eqref{eq:7:language} as shown below, whereby we distinguish between two kinds of violations, \ie return values $z$ that are less than $x+1$ and return values that are greater than $x+1$, and require the latter violations to be detected in synchronous fashion.\\
  \begin{minipage}[c]{0.90\linewidth}
   \begin{equation*}
    \toolMax{\hVarX}{\toolNec{\toolActIn{\toolN{server}}{\toolTup{\toolN{succ},x,y}}}{\;\toolNec{\toolActOut{y}{z}}{\;
          \left(
            \begin{array}{l}
              {(\toolBool{(z\!=\!x\!+\!1)}{\hVarX\,})} \\
              \quad \toolAnd{}{
                \toolAnd{\;
                  (\toolBool{(z< x\!+\!1)}{\toolFls\,})
                \\\quad}{\;\;
                  (\toolBool{(z > x\!+\!1)}{\toolSFls\,})
                }}
            \end{array}
          \right)
        }}} 
  \end{equation*}
\end{minipage}
\begin{minipage}[l]{0.09\linewidth}
\vspace{1.3cm}
  \qedd
\end{minipage}
\end{example}

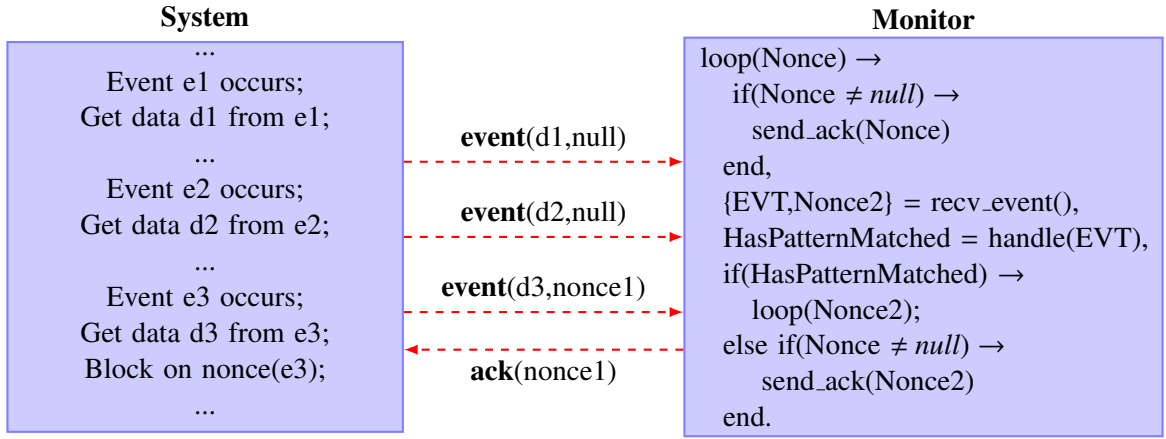
\begin{figure}[t]
  \centering
  \begin{tikzpicture}[>=latex,auto,thick]
    \begin{scope}[draw=blue!50,fill=blue!20,minimum size=0.8cm]
      \node (sys) at ( -1,1) [shape=rectangle,draw,fill,text width=5cm, minimum height = 3cm, text centered] 
          {...\\ Event e1 occurs;\\ Get data d1 from e1;\\...\\Event e2 occurs;\\ Get data d2 from e2;\\...\\Event e3 occurs;\\ Get data d3 from e3;\\Block on nonce(e3);\\...\\\; };
       \node (mon) at ( 8.55,1) [shape=rectangle,draw,fill,text width=6.1cm, minimum height = 3cm] {$\; $loop(Nonce)\;$\rightarrow$\\$\quad$ if(Nonce $\not{=} null$)\;$\rightarrow$\\ $\qquad$send\_ack(Nonce) \\ $\quad$end, \\ $\quad$\{EVT,Nonce2\} = recv\_event(), \\ $\quad$HasPatternMatched = handle(EVT), \\ $\quad$if(HasPatternMatched) $\rightarrow$ \\ $\qquad$loop(Nonce2); \\ $\quad$else if(Nonce $\not{=} null$)\;$\rightarrow$\\ $\qquad$ send\_ack(Nonce2)\\ $\quad$end. };  
        \node (sysN) at (-1,3.9) {\textbf{System}}; 
        \node (monN) at (8.55,3.9) {\textbf{Monitor}}; 
    \end{scope}
       \node (sys1) at (1.5,2.0){};
       \node (mon1) at (5.5,2.0){};
       \node (sys2) at (1.5,1.0){};
       \node (mon2) at (5.5,1.0){};
       \node (sys3) at (1.5,0){};
       \node (mon3) at (5.5,0){};
       \node (sys4) at (1.5,-0.5){};
       \node (mon4) at (5.5,-0.5){};
    \begin{scope}[draw=red,fill=red, dashed]
      \draw[->] (sys1) to node{\textbf{event}(d1,null)} (mon1); 
      \draw[->] (sys2) to node{\textbf{event}(d2,null)} (mon2); 
      \draw[->] (sys3) to node{\textbf{event}(d3,nonce1)} (mon3); 
      \draw[->] (mon4) to node{\textbf{ack}(nonce1)} (sys4); 
    \end{scope}
  \end{tikzpicture}
  \caption{A high-level depiction of the hybrid monitoring protocol. 
  }
       \label{fig:hybrid_protocol}
\end{figure}

The new monitor synthesis algorithm  requires a pre-processing phase to determine which events are to be synchronously monitored in order to implement a synchronous fail. For instance, formulas $\toolNec{\alpha}{\toolSFls}$ and $\toolSNec{\alpha}{\toolFls}$ are both monitored in the same way, in fact the pre-processing phase encodes the former into the latter. In general, however, determining which actions to synchronously monitor for implementing a synchronous fail is not as straightforward, since the $\toolSFls$ and the first necessity formula preceding this $\toolSFls$ may be interposed by intermediate formulas such as conjunctions and boolean guards (as in the property of \exref{ex:synch-asynch}). In such a case the compiler inspects each boolean guard and checks whether there exists at least \emph{one} boolean guard that leads directly to a synchronous fail, i.e., {\toolBool{b}{\toolSFls}}, and if so, the action specified in the necessity is synchronously monitored. In hybrid monitoring, both synchronous and asynchronous event monitoring require code instrumentation, as shown in \figref{fig:hybrid_protocol}.  \emph{Asynchronous} events inject advice functions sending a monitor message 
containing the event details and a \emph{null} nonce, \emph{without blocking} the system; upon receiving a null nonce, the monitor determines that it does not need to send an acknowledgment back to the system. \emph{Synchronous} actions are implemented as before (see \secref{fig:sync_protocol}), where \emph{fresh} nonces indicates the monitor that it needs to acknowledge back.


\section{Evaluation}
\label{sec:evaluation}

\begin{figure}
\centering
	\includegraphics[width=0.95\textwidth]{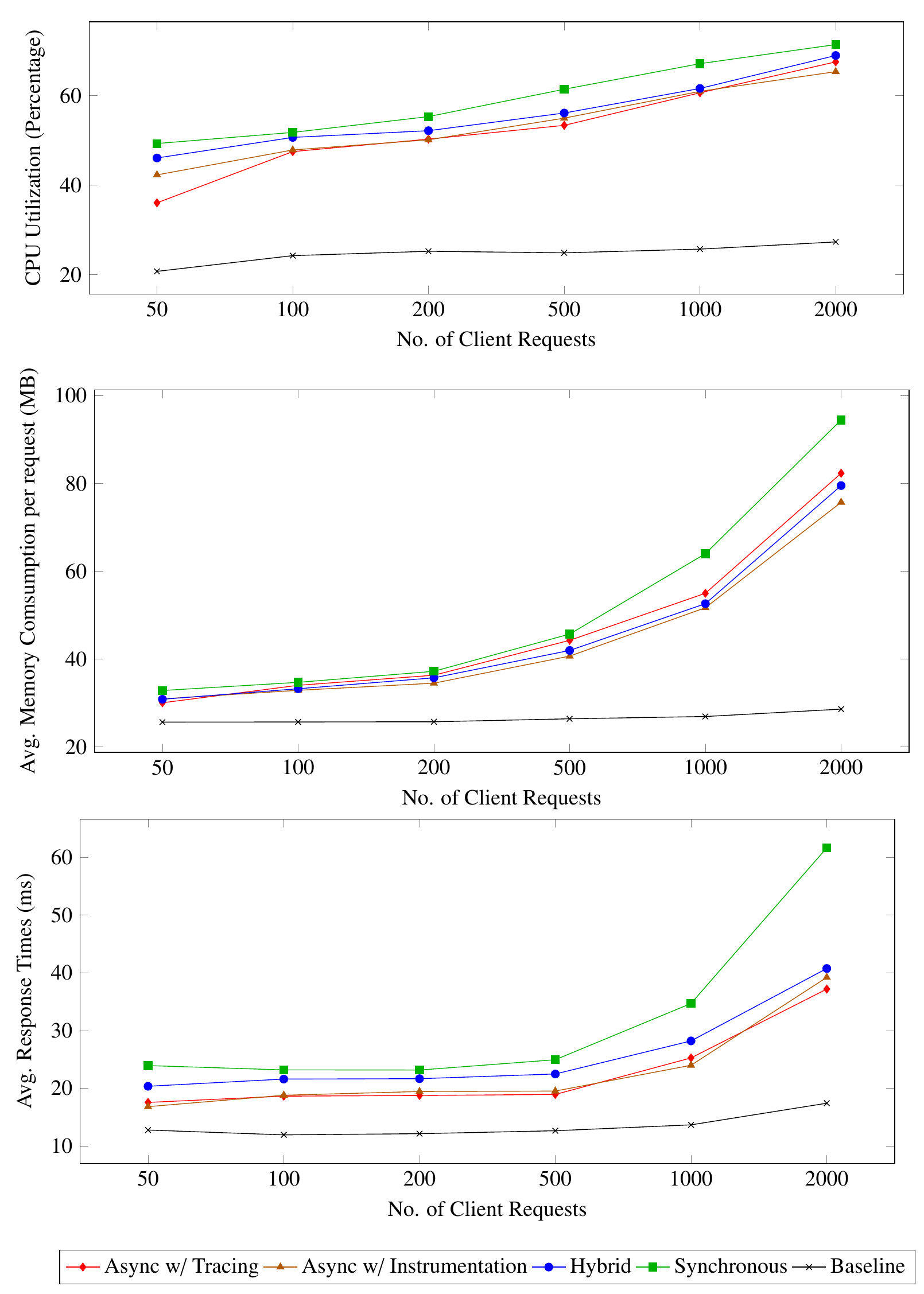}
	\caption{These graphs denote the average cpu time, memory utilization and response time when monitoring the system using different monitoring approaches.}
	\label{fig:combined_results}
\end{figure}

Using the extended syntax, we reformulate the security properties used for the evaluation of \secref{sec:preliminary-results} and require violation detections to be synchronous \ie using \toolSFls instead of \toolFls.  For instance, from \eqref{eq:5} we obtain the  property below.
\begin{equation}
\begin{array}{l}
    \toolMax{\hVarX}{\\\quad\toolSNec{\toolActOut{\toolN{AcceptorPid}}{\toolTup{handlerPid,\toolN{next},\_}}}\\
      \quad\toolAnd{\Bigl( X\;}{}
      \;\toolMax{\hVarY}{\toolNec{\toolActRet{handlerPid}{\toolTup{\toolN{yaws}, \toolN{do\_recv},\toolN{3},\toolTup{\toolN{ok},\toolTup{\toolN{http\_req},\toolN{GET},\_ ,\_}}}}}\\
    \qquad\toolNec{\toolActRet{handlerPid}{\toolTup{\toolN{yaws}, \toolN{do\_recv},\toolN{3},\toolTup{\toolN{ok},h1}}}}\\
    \qquad \vdots \\
    \qquad\toolNec{\toolActRet{handlerPid}{\toolTup{\toolN{yaws},\toolN{do\_recv},\toolN{3},\toolTup{\toolN{ok},h6}}}}\\
    \qquad
            \left(\begin{array}{l}
            (\toolBool{(\textsl{isMalicious}(h1,h2,h3,h4,h5,h6))}{\;\;\toolSFls\;})\\
            \toolAnd{}{\bigl(\toolBool{(\lnot \textsl{isMalicious}(h1,h2,h3,h4,h5,h6))}{}\\
            \qquad\qquad\toolNec{\toolActRet{handlerPid}{\toolTup{\toolN{yaws},\toolN{do\_recv},\toolN{3},\toolTup{\toolN{ok},\toolN{http\_eoh}}}}}{\hVarY})}\bigr)
          \end{array}
          \right)\\
     \quad\Bigr) }}
\end{array}\label{eq:6}
\end{equation}
We measure the respective overheads resulting from the hybrid instrumentations over Yaws for varying client loads  in terms of $(i)$ the average CPU utilization required; $(ii)$ the memory overheads per Yaws client request; and $(iii)$ the average time taken for the (monitored) Yaws server to respond to batches of simultaneous client request.\footnote{In the extended syntax, properties that are exclusively defined in terms of synchronous necessity formulas, \toolSNec{\actE}{\hV}, yield synchronous monitor instrumentations that are identical to those discussed in \secref{sec:synchr-instr}.}  The experiments were carried out on an Intel Core 2 Duo T6600 processor with 4GB of RAM, running Microsoft Windows 7 and EVM version R16B03.  For each property and each client load, we take three sets of readings and then average them out.  Since results do not show substantial variations when different properties were considered, we again average them across all properties and compile them in the graphs shown in \figref{fig:combined_results}.   

The results show that the hybrid instrumentation yielded CPU utilisation and memory overheads that are substantially lower than those incurred by a synchronous instrumentation, 
comparable to those 
of asynchronous monitoring with code injections.   The second graph even shows that the memory utilisation for both of these instrumentations is less than that for asynchronous monitoring performed through the EVM tracing of \cite{Fra:Sey:13}.    In the case of response-time latencies, where synchronous monitoring fared the worst  (on average 30\% higher overheads than its asynchronous counterpart) a hybrid approach managed to lower response times to overheads that are about 15\% higher than asynchronous monitoring; see \figref{fig:combined_results} bottom graph.

%% file: conc.tex
We studied various monitoring techniques for actor-based frameworks in the context of Erlang and integrated them within a tool for runtime verification of actor systems.  Our contributions are:
\begin{itemize}
\item A novel hybrid instrumentation technique minimising the amount of (expensive) synchronous monitor instrumentations while still guaranteeing timely violation detections.
\item An extension to \detecter, an RV tool for Erlang (actor-based) programs, that allows the verifier to control which violations to monitor synchronously and asynchronously within the same property.
\item A case study demonstrating the applicability of the technique and tool to monitor safety properties for Yaws, a concurrent web-server written in Erlang.
\item A systematic assessment of the relative overheads incurred by different instrumentation techniques within an actor setting. 
  
\end{itemize}

\paragraph{Related Work:}
\label{sec:rel-work}

Several verification and modeling tools \cite{marjan2004,fredlund:mcerlang,maude,rt-synchronizer,EfficientDec} for actor-based component systems already exist. Rebeca \cite{marjan2004,marjan2011} is an actor-based modeling language designed with the aim of bridging the gap between formal verification approaches and real applications. It provides conversions to renowned model checkers like SMV and Promela, and extends the model with timing constraints; timed-rebeca models have also been translated into Erlang.  McErlang \cite{fredlund:mcerlang} is a model-checker specifically targeting Erlang code; it uses a superset of our logic.  We are however unaware of any extensions of these tools to RV.  Apart from \detecter \cite{Fra:Sey:13},  other RV tools for actor based system exist. \elarva \cite{elarva:2012} is another Erlang monitoring tool that  uses the EVM tracing mechanism to perform \emph{asynchronous} monitoring;  no facility for synchronous monitoring is provided.  In \cite{EfficientDec}, Sen \etal explore a decentralized (orchestrated) monitoring approach as a way to reduce the communication overheads that are usually caused by a centralized approach and implement it in terms of an actor-based tool called \textsc{DiAna}. Although their investigation is orthogonal to ours, it would be interesting to integrate this study within ours and systematically evaluate whether choreographed monitor arrangements yield further overhead gains.

By and large, most widely used online RV tools employ synchronous instrumentation \cite{java-mac,larva-CPS09,chen-rosu-2007-oopsla, jUnitRv:13,BarringerFHRR12}. There is also a substantial body of work commonly referred to as asynchronous RV \cite{lola:runtime,d'Amorim:2005:ERV:1082983.1083249,Andrews03generaltest}. However, the latter tools and algorithms assume \emph{completed traces}, generated by complete program executions and stored in a database or a log file;  as explained in the Introduction, we term these bodies of work as \emph{offline}, and their aims are considerably different from the work presented here. There exist a few tools offering both synchronous and asynchronous monitoring, such as MOP \cite{chen-rosu-2005-tacas,chen-rosu-2007-oopsla} and JPAX \cite{Havelund:2001:MJP:891177,SyncVSAsync:Rosu:2005}.  Crucially, however, they do not provide the fine grained facility of supporting both synchronous and asynchronous monitoring at the same time, or switching between the two modes at runtime. 

Rosu \etal \cite{SyncVSAsync:Rosu:2005}  make a similar distinction to ours between offline, synchronous online  and asynchronous online monitoring. However, their definitions of synchrony and asynchrony deals with the timeliness of detections. By contrast  we focus on how instrumentation is carried out and show how hybrid instrumentation can be used to obtain timely detections for certain properties; this amounts to synchronous monitoring in \cite{SyncVSAsync:Rosu:2005}. A closer work to ours is  \cite{CP12FF}: they allow a decoupling between the system and monitor executions but provide explicit mechanisms for pausing the system while the lagging 
monitor execution catches up. In our case, this mechanism is handled \emph{implicitly}  when switching from asynchronous to synchronous monitoring.  In \cite{CP12FF} they do not provide an implementation of their constructs and do not asses 
the relative overhead costs incurred by different 
instrumentation strategies. 

 Talcott \etal in \cite{Talcott2008} compare and contrast three coordination models for actors which cover a wide spectrum of communication mechanisms and coordination strategies. 
 The comparison focusses on  the level of expressivity of each model, the level of maturity, 
 the level of abstraction 
 and the way user definable coordination behavior is provided. One of the analysed coordination models, ie. the Reo model, is a channel based language in which channels may be either synchronous or asynchronous. This model resembles the way our hybrid monitoring protocol interacts with the monitored system. In fact our monitoring language constructs \texttt{[$\alpha$]} and \texttt{[|$\alpha$|]}, seem analogous to the asynchronous and synchronous channels in the Reo model.  


\paragraph{Future Work:}
\label{sec:future-work}
Our hybrid instrumentation technique can be extended to other inherently asynchronous computational models, such as monitoring for distributed systems \cite{FGP12DistribRV}.  This would require us to consider additional aspects such as the handling of multiple, partially-ordered traces and the use of alternative monitor organisations such as monitor choreographies.  Our techniques can also be extended with enforcement mechanism \cite{FalconeFM12}, facilitated by the fact that corrective action can be carried out  as soon as the system performs the violation.  This would be worth exploring in the context of Erlang and \detecter.  



